\documentclass[epj]{svjour}
\usepackage{xcolor}
\usepackage{graphics}
\begin{document}

\title{Neutron star equation of state and tidal deformability with nuclear energy density functionals}
\author{Young-Min Kim\inst{1}, Kyujin Kwak\inst{1} \and Chang Ho Hyun\inst{2} \and 
Hana Gil\inst{3} \and Chang-Hwan Lee\inst{4}% etc
\thanks{\emph{E-mail:} clee@pusan.ac.kr}%
}                     % Do not remove

\institute{School of Natural Science, Ulsan National Institute of Science and Technology (UNIST), Ulsan 44919, Korea 
\and Department of Physics Education, Daegu University, Gyeongsan 38453, Korea
\and Department of Physics, Kyungpook National University, Daegu 41566, Korea
\and Department of Physics, Pusan National University, Busan 46241, Korea}
\date{Received: date / Revised version: date}

\abstract{
Neutron star is the ultimate testing place for the physics of dense nuclear matter. Before the detection of gravitational waves from the merger of binary neutron stars, various nuclear equations of state have been used to estimate the macroscopic properties of neutron stars, such as masses and radii, based on the electromagnetic observations.
However, recent observations on the tidal deformability of neutron star from the gravitational waves GW170817 opened a new era of multi-messenger astronomy and astrophysics, and many theoretical works have been extended to estimate the tidal deformability of neutron stars. 
In this article, we review our recent works on the application of nuclear energy density functionals to the properties of neutron stars including tidal deformability. We found that many nuclear energy density functionals, including new KIDS (Korea: IBS-Daegu-Sungkyunkwan)  model, satisfy both constraints from current electromagnetic and gravitational wave observations. We discuss future possibilities of constraining nuclear matter equation of state from ground-based experiments and multi-messenger observations.
\PACS{
      {PACS-key}{neutron star equation of state} \and {PACS-key}{tidal deformability}   \and
      {PACS-key}{energy density functional}
     } % end of PACS codes
} %end of abstract

\authorrunning{Kim et al.}
\titlerunning{NS EOS and tidal deformability with nuclear energy density functionals}

\maketitle

%--------------------------------------------------------------------------------------------------
\section{Introduction} 
\label{intro}

Neutron star is a compact stellar object which can provide valuable information on the physics of dense nuclear matter because the density of the central part of a neutron star can reach several times of the normal nuclear matter density. In this dense environment, the quantum degeneracy pressure of hadrons dominates and the microscopic calculation based on QCD is almost impossible due to the highly non-perturbative nature of the strong interaction. Hence many effective approaches including energy density functionals have been used to understand the properties of nuclear matter and the macroscopic quantities of neutron stars such as masses and radii.

Most of the observed neutron stars are in radio pulsars. Even though  the magnetic field strength of most pulsars can be measured, 
if they are not in close binaries, their masses and radii are not easily extractable. Masses and radii of neutron stars have been measured in various kinds of binaries; low-mass X-ray and optical binaries, neutron star-white dwarf binaries, double neutron star binaries, and neutron star-main sequence binaries~\cite{Pra13}. In two of neutron star-white dwarf binaries,  $2 M_\odot$ neutron stars have been observed~\cite{demorest2010,antoniadis2013,cromartie19}. These observations set the maximum mass of neutron stars to lie above $2 M_\odot$. On the other hand, most of the well-measured neutron star masses in double neutron star binaries lie below $1.5 M_\odot$. These observations imply that the actual masses of neutron stars may strongly depend on the evolution of neutron star binaries and the evolution of their progenitors~\cite{Lee14,Lee17}.

Detection of gravitational waves (GW170817) followed by gamma-ray burst (GRB 170817A) and electromagnetic afterglows (AT 2017gfo) from the merger of a neutron star binary opened a new era of multi-messenger astronomy and astrophysics~\cite{GW170817PRL,GW170817ApJ}. First of all, the estimation of the masses and the tidal deformabilities of neutron stars from gravitational wave signals in GW170817 proved that gravitational waves can provide valuable information on the inner structure of neutron stars. This implies that nuclear and particle physics in dense environments can be tested by gravitational waves. In this article, we discuss the neutron star equation of state based on the nuclear energy density functionals. Secondly, series of afterglow observations in various frequency bands confirmed a possibility that a kilonova can be formed by a merger of neutron stars in a binary. Since the mergers of neutron star binaries can provide neutron rich environment which is essential for the production of heavy elements via r-process, it is believed that many heavy elements are produced in this kilonova.

As noticed above, multi-messenger observations of the merger of a neutron star binary provided new possibilities to nuclear and particle physics. In the work of Kim et al.~\cite{Kim2018a,Kim2018b}, we investigated several Skyrme forces and energy density functionals (EDFs) including KIDS (Korea: IBS-Daegu-Sungkyunkwan) which are consistent with the properties of nuclear matter in finite nuclei and the constraints from neutron star observations. KIDS model is a density functional model which has been developed in Korea~\cite{KIDSprc97}. 
In this article, we review the expansion scheme of energy density functional KIDS and its applications to finite nuclei and neutron stars.
In Sec.~\ref{sec2}, we summarize KIDS and EDFs used in Kim et al.~\cite{Kim2018a,Kim2018b}, focusing on the constraints from various nuclear physics experiments. In Sec.~\ref{sec3}, we summarize our numerical results on the neutron star properties, focusing on the constraints provided by the electromagnetic and gravitational wave observations. In Sec.~\ref{sec4}, we discuss the future possibilities of multi-messenger observations and ground-based experiments.

%--------------------------------------------------------------------------------------------------
\section{Nuclear matter properties with nuclear energy density functionals} 
\label{sec2}

Neutron stars provide conditions extreme both in density $\rho = \rho_n + \rho_p$ and in the neutron-proton asymmetry $\delta = (\rho_n-\rho_p)/(\rho_n+\rho_p)$ where $\rho_n$ and $\rho_p$ are the neutron and the proton density, respectively. An important issue in the recent study of nuclear structure and dense nuclear matter theory is to understand the validity of predictions obtained from extrapolation to these extreme conditions, and to make the quantitative estimation of uncertainty feasible. Among several possibilities, an expansion scheme with a suitable expansion variable can provide a way to realize the uncertainty quantification for nuclei and neutron stars.

The uncertainty quantification constitutes an essential part of the motivation to develop the KIDS EDF model, which aims to construct a theoretical frame that can provide reliable extrapolation and prediction in the extreme conditions of neutron stars. Based on the observation that $k_F/m_\rho$ could be an expansion parameter for the energy density of infinite nuclear matter relevant to neutron stars,
where $k_F$ is the Fermi momentum and $m_\rho$ is the rho-meson mass, we assume expansion of the energy per particle in homogeneous nuclear matter in the powers of the cubic root of the density as
\begin{eqnarray}
{\cal E} (\rho, \delta) = {\cal T}(\rho, \delta) + \sum_{n=0}^{N-1} (\alpha_n + \beta_n \delta^2) \rho^{1+n/3},
\end{eqnarray} 
where ${\cal T}$ is the kinetic energy, and the summation represents the contribution from strong interactions. We assume quadratic approximation for the asymmetry $\delta$. Parameters $\alpha_n$ and $\beta_n$ are fit to empirical or well-known properties of symmetric and asymmetric nuclear matters, respectively. It has been shown that the number of terms necessary for an optimal description of the nuclear matter in the density range from well below the saturation to the core of the heaviest neutron star is 7, which consists of 3 from the symmetric part ($\alpha_0 \sim \alpha_2$) and 4 from the asymmetric part ($\beta_0 \sim \beta_3$) \cite{KIDSprc97}. In addition, it is shown that even if we consider terms more than 3 for the symmetric matter and 4 for the asymmetric matter, additional terms seldom affect the predictions for the basic properties of nuclei and neutron stars \cite{prc100}. Small difference in the results with and without additional terms suggests the amount of uncertainty from higher order contributions.

Nuclear matter is an extension of finite nuclei in the limit of infinite number of nucleons. In a majority of cases, models of nuclear structures are determined by fitting the parameters to a set of selected nuclear data, and the nuclear matter properties are obtained as results (or predictions) of the model. Compared to this conventional approach, our approach with KIDS model takes a reverse engineering: model parameters are fit to the nuclear matter equation of state first, and additional parameters in the following steps are determined from nuclear data. Extention of KIDS model to the application to nuclei is described in depth in \cite{prc99}. We note that in any case, regardless of the fitting procedure, a realistic nuclear structure model should describe both finite nuclei and infinite nuclear matter well and simultaneously.

\begin{figure*}
\begin{center}
\resizebox{0.8\textwidth}{!}{
\includegraphics{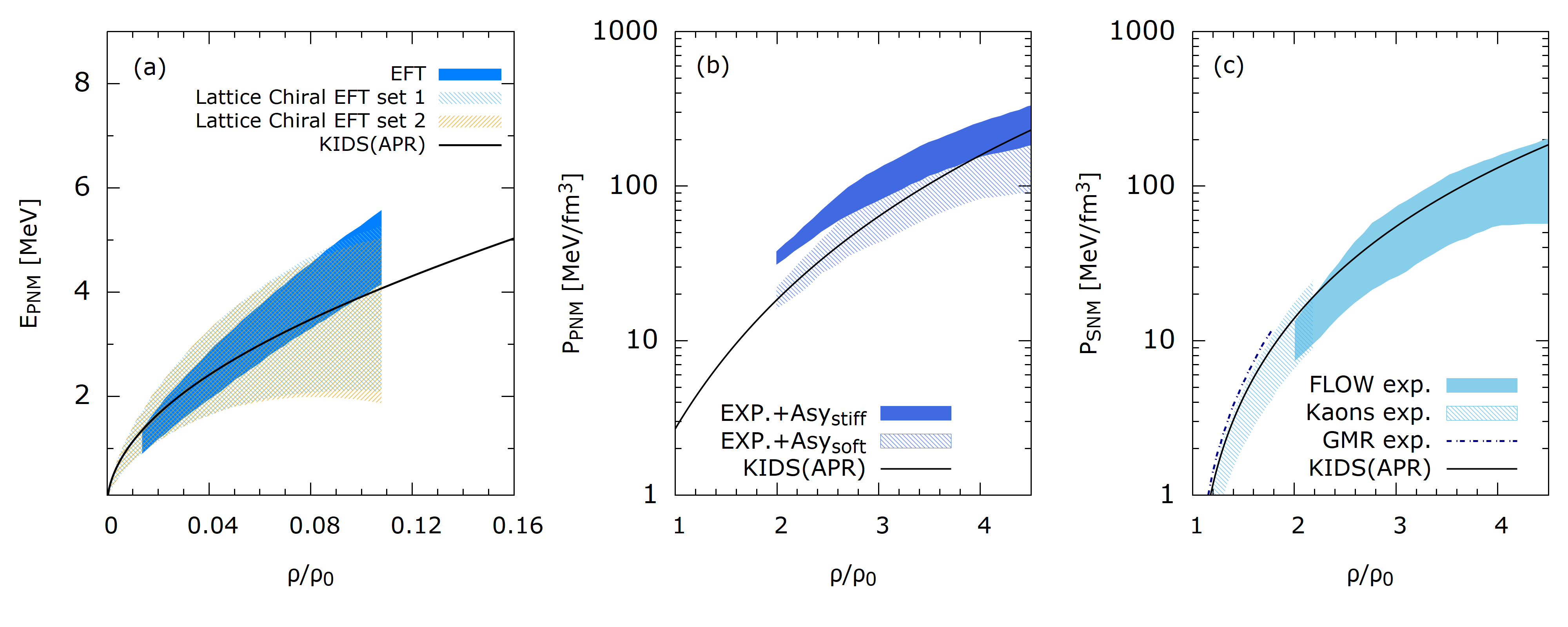}
}
\end{center}
\caption{Energy below (a) and pressure above (b, c) the saturation density.
Result of an EFT and lattice chiral EFT in (a) is taken from \cite{dss} and \cite{lattice}, respectively. 
Results from experiment in (b) are from \cite{flow}, and those in (c) are from \cite{flow,kaon,Fuc06,You99}.
}
\label{fig1}       
\end{figure*}

In this work, we adopt the parameterization denoted as `KIDS-ad2' in \cite{KIDSprc97}. In the KIDS-ad2 model, $\alpha_0 \sim \alpha_2$ are determined to produce three basic properties of symmetric nuclear matter, saturation density $\rho_0$ ($=0.16$ fm$^{-3}$), binding energy per nucleon $E/A$ ($=16$ MeV), and incompressibility $K_0$ ($=240$ MeV) at the saturation density. With thus determined $\alpha_n$'s, $\beta_0 \sim \beta_3$ are fit to the equation of state of pure neutron matter calculated theoretically \cite{apr}. In Ref.~\cite{dutra}, 11 experimental/empirical (Exp/Emp in short) data about the nuclear matter properties are adopted to test nuclear models.
Among the 240 Skyrme force models, only 16 models satisfy all the 11 Exp/Emp constraints. In Tab. \ref{tab1} and Fig. \ref{fig1}, we show the nuclear matter properties obtained from the KIDS model and compare them with Exp/Emp ranges.

In Table \ref{tab1}, $\rho_0$, $E/A$ and $K_0$ are assumed to be the given values, and the parameters of the KIDS model in the symmetric part are solved to reproduce them. $Q_0$ is obtained as a consequence of the solution. On the other hand, since the parameters in the asymmetric part are fitted to the pure neutron matter (PNM) equation of state in \cite{apr}, symmetry energy parameters $J$, $L$ and $K_\tau$ are calculated as results of the fitting. Figure \ref{fig1}(a) shows the energy of PNM at sub saturation densities. Theoretical equations of state calculated with an EFT \cite{dss} and a lattice chiral EFT \cite{lattice} are consistent with the prediction of KIDS model.
In the subsequent works, contributions from the chiral three-nucleon forces are taken into account \cite{Dri16,Tew16}.  Around $\rho \sim 0.016-0.020$ fm$^{-3}$ ($=0.1-0.125 \rho_0$), new calculations predict $E_{\rm NM} \sim 4$ MeV, so the result of KIDS model is consistent with the updated predictions from EFTs.
One can also find $E_{\rm PNM}$ at very low densities calculated with Av4 potential and quantum Monte Carlo method \cite{qmcav4} (QMC Av4). 
In Ref. \cite{prc99}, $E_{\rm PNM}/E_{\rm FG}$ ($E_{\rm FG}$ the energy of the free gas) is compared with various models.
Though not fully consistent, KIDS model prediction agrees to the result of QMC Av4 in the dilute neutron matter. Figure \ref{fig1}(b, c) presents the supra saturation-density behavior of pressure in PNM (b), and in the symmetric nuclear matter (SNM) (c). Model prediction is compared to the result from heavy ion collision experiments.

%Table1 
\begin{table*}
\caption{Properties of nuclear matter calculated with KIDS model. Saturation density $\rho_0$ is in unit of fm$^{-3}$. Exp/Emp values are quoted from Ref.~\cite{dutra}. $E/A$, $K_0$, and $Q_0$ are binding energy per particle, compression modulus, and skewness (the third derivative of energy per particle) at the saturation density in the symmetric nuclear matter, respectively. $J$, $L$, and $K_\tau$ are parameters in the symmetry energy of nuclear matter. $E_0$, $K_0$, $Q_0$, $J$, $L$, and $K_\tau$ are in units of MeV.}
\label{tab1}    
\begin{center}
\begin{tabular}{llllllll}
\hline\noalign{\smallskip}
 & $\rho_0$ & $E/A$ & $K_0$ & $-Q_0$, & $J$ & $L$ & $-K_\tau$  \\
\noalign{\smallskip}\hline\noalign{\smallskip}
KIDS & $0.160$ & $16.00$ & $240.0$ & $372.7$ & $32.8$ & $49.1$ & $375.1$ \\
Exp/Emp & $\simeq 0.16$ & $\simeq 16.0$ & 200-260 & 200-1200 & 30-35 & 40-76 & 372-760 \\
\noalign{\smallskip}\hline
\end{tabular}
\end{center}
\end{table*}

%Fig1

As aforesaid, a realistic nuclear structure model should be able to reproduce basic properties of nuclei. We transform the KIDS EDF to a Skyrme-type in-medium nuclear potential, and solve the Hartree-Fock equation to obtain the wave function of a nucleus \cite{prc100,prc99}.
Two model parameters are added when a Skyrme-type potential is constructed. One term accounts for the gradient of density distribution in a nucleus, and the other for the spin-orbit interactions. They are fit to the binding energy per nucleon and charge radius of $^{40}$Ca, $^{48}$Ca, and $^{208}$Pb, and the rest of the nuclear properties are obtained as predictions of the model. Tables \ref{tab2} and \ref{tab3} show the result of KIDS model and compare them with experimental data. Though only small portion of parameters (2 among 9) is fit to nuclear data, prediction of the bulk properties of spherical magic nuclei agrees to experiment remarkably well with deviation at the order of 0.1\% or less for medium and heavy nuclei.

%Table 2
\begin{table*}
\caption{Binding energy per nucleon in units of MeV. Deviation is defined as $|{\rm Exp}-{\rm KIDS}|/{\rm Exp}\times 100$.  Data are taken from \cite{exp1,exp2}. }
\label{tab2}    
\begin{center}
\begin{tabular}{ccccccc}
\hline\noalign{\smallskip}
 & $^{16}$O &  $^{40}$Ca & $^{48}$Ca & $^{90}$Zr & $^{132}$Sn & $^{208}$Pb  \\
\noalign{\smallskip}\hline\noalign{\smallskip}
Exp & 7.9762 & 8.5513 & 8.6667 & 8.7100 & 8.3550 & 7.8675 \\
KIDS & 7.8684 &  8.5565  & 8.6564 &  8.7328 & 8.3563 & 7.8809 \\
Deviation (\%) & 1.35 &  0.06 & 0.12 & 0.26 & 0.02 & 0.17 \\
\noalign{\smallskip}\hline
\end{tabular}
\end{center}
\end{table*}

%Table 3
\begin{table*}
\caption{Charge radius in units of fm.
Deviation is defined as $|{\rm Exp}-{\rm KIDS}|/{\rm Exp}\times 100$.  Data are taken from \cite{exp1,exp2}. }
\label{tab3}    
\begin{center}
\begin{tabular}{ccccccc}
\hline\noalign{\smallskip}
 & $^{16}$O &  $^{40}$Ca & $^{48}$Ca & $^{90}$Zr & $^{132}$Sn & $^{208}$Pb  \\
\noalign{\smallskip}\hline\noalign{\smallskip}
Exp & 2.6991 & 3.4776 & 3.4771 & 4.2694 & 4.7093 & 5.5012 \\
KIDS & 2.7618 & 3.4781 & 3.4867 & 4.2476 & 4.7089 & 5.4887 \\
Deviation (\%) & 2.32 & 0.01 & 0.28 & 0.51 & 0.01 & 0.23 \\
\noalign{\smallskip}\hline
\end{tabular}
\end{center}
\end{table*}

It has been shown in many papers that nuclear equation of state calibrated to similar conditions at the saturation density can behave very differently at high densities. Masses of the neutron stars estimated from the observation GW170817 are about $1.4 M_\odot$. Density at the center of $1.4 M_\odot$ neutron star is in the range $(2 \sim 3) \rho_0$, the uncertainty being originated from the nuclear model. In order to study of the uncertainty of the tidal deformability due to the dependence on the nuclear model, we consider four non-relativistic Skyrme force models GSkI \cite{gsk1}, SLy4 \cite{sly4}, SkI4 \cite{ski4}, SGI \cite{sg1}, and two relativistic mean field models MS1, MS1b \cite{ms1}. Table \ref{tab4} shows that the models have similar nuclear matter properties at the saturation density, and they are also in reasonable agreement with experiment.

%Table4
\begin{table*}
\caption{Four basic nuclear matter properties obtained from models in comparison. Saturation density $\rho_0$ is in fm$^{-3}$, and $E/A$, $K_0$ and $J$ are in units of MeV. MS1b model gives the results same with MS1.}
\label{tab4}    
\begin{center}
\begin{tabular}{cccccc}
\hline\noalign{\smallskip}
 & GSkI &  SLy4 & SkI4 & SGI & MS1  \\
\noalign{\smallskip}\hline\noalign{\smallskip}
$\rho_0$ & 0.159 & 0.160 & 0.160 & 0.154 & 0.148 \\
$E/A    $ & 16.02 & 15.97 & 15.95 & 15.89 & 15.75  \\
$K_0     $ & 230.2 & 229.9 & 248.0 & 261.8 & 250 \\
$J        $ & 32.0 & 32.0 & 29.5 & 28.3 & 35 \\
\noalign{\smallskip}\hline
\end{tabular}
\end{center}
\end{table*}

%Figure2
\begin{figure}[t]
\begin{center}
\resizebox{0.4\textwidth}{!}{\includegraphics{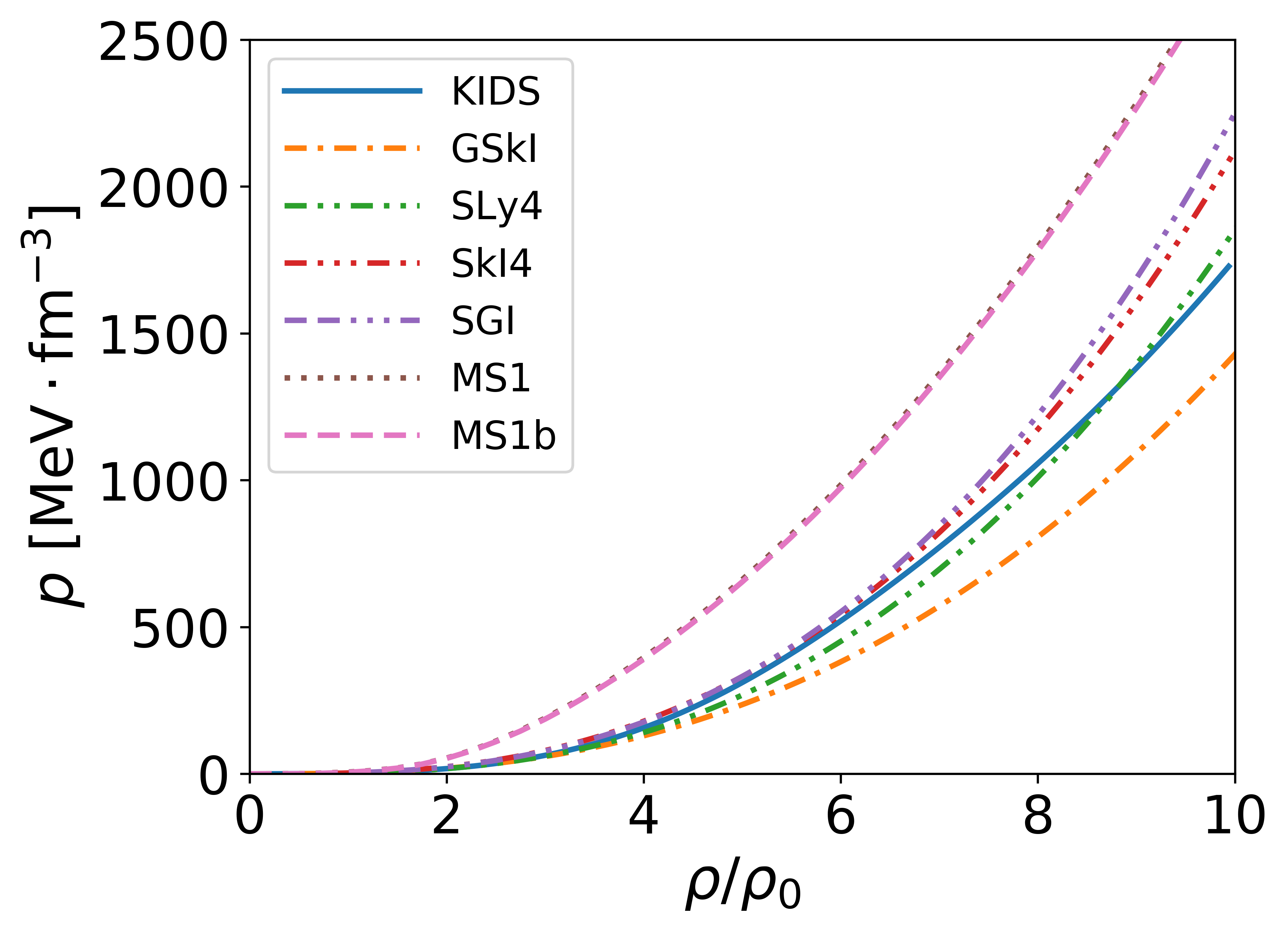}}
\centerline{(a) $p$ vs $\rho/\rho_0$}
\resizebox{0.4\textwidth}{!}{\includegraphics{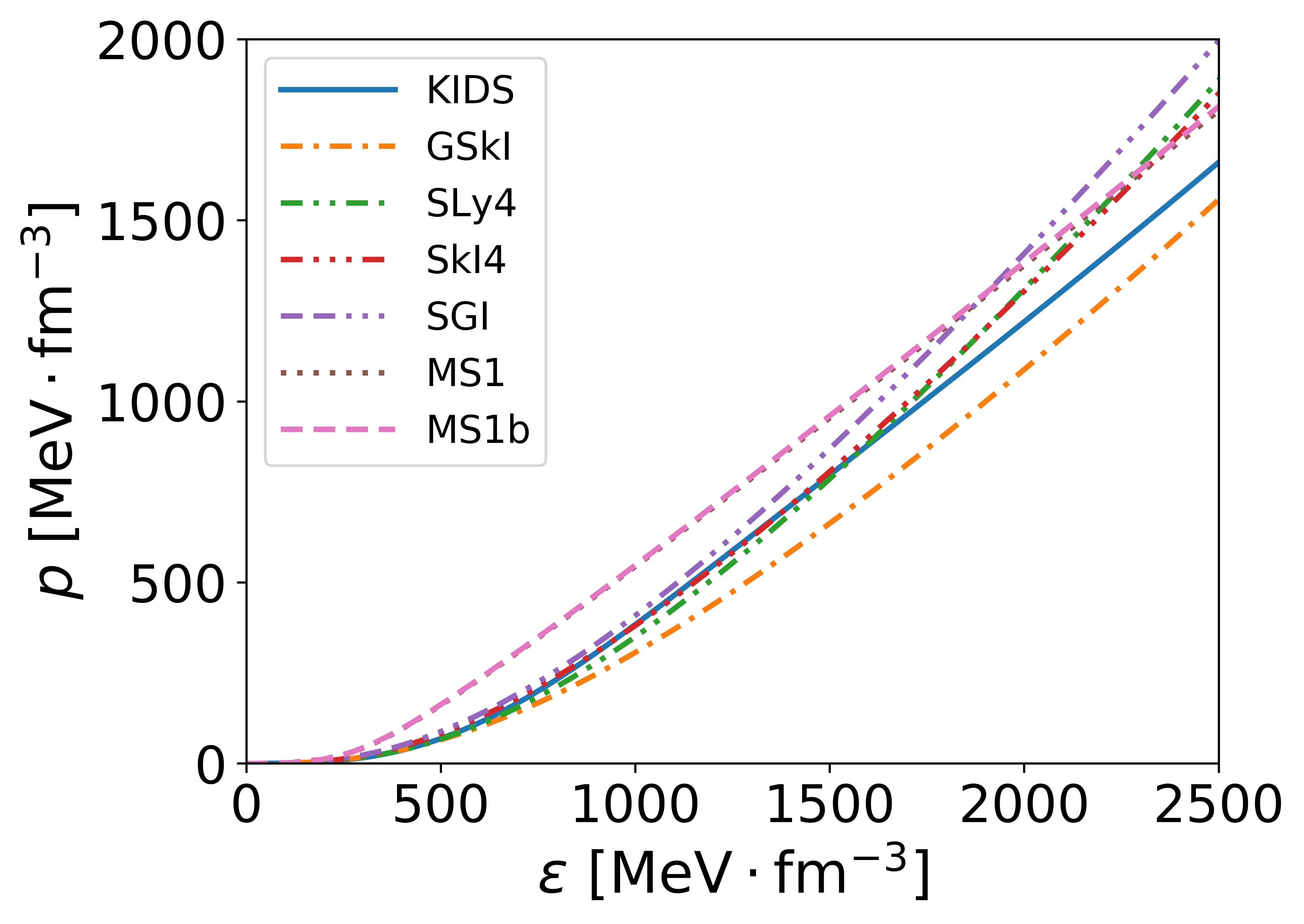}}
\centerline{(b) $p$ vs $\epsilon$}
% \resizebox{0.49\textwidth}{!}{\includegraphics{fig1-ep-cgs.png}}
\end{center}
\caption{Pressure ($p$) vs density ($\rho/\rho_0$) and energy density ($\epsilon$) for various nuclear equations of state~\cite{Kim2018a,Kim2018b}.}
\label{fig2}
\end{figure}

%--------------------------------------------------------------------------------------------------
\section{Neutron star equation of state and tidal deformability with nuclear energy density functionals}
\label{sec3}

\begin{figure}[t]
\begin{center}
\resizebox{0.4\textwidth}{!}{\includegraphics{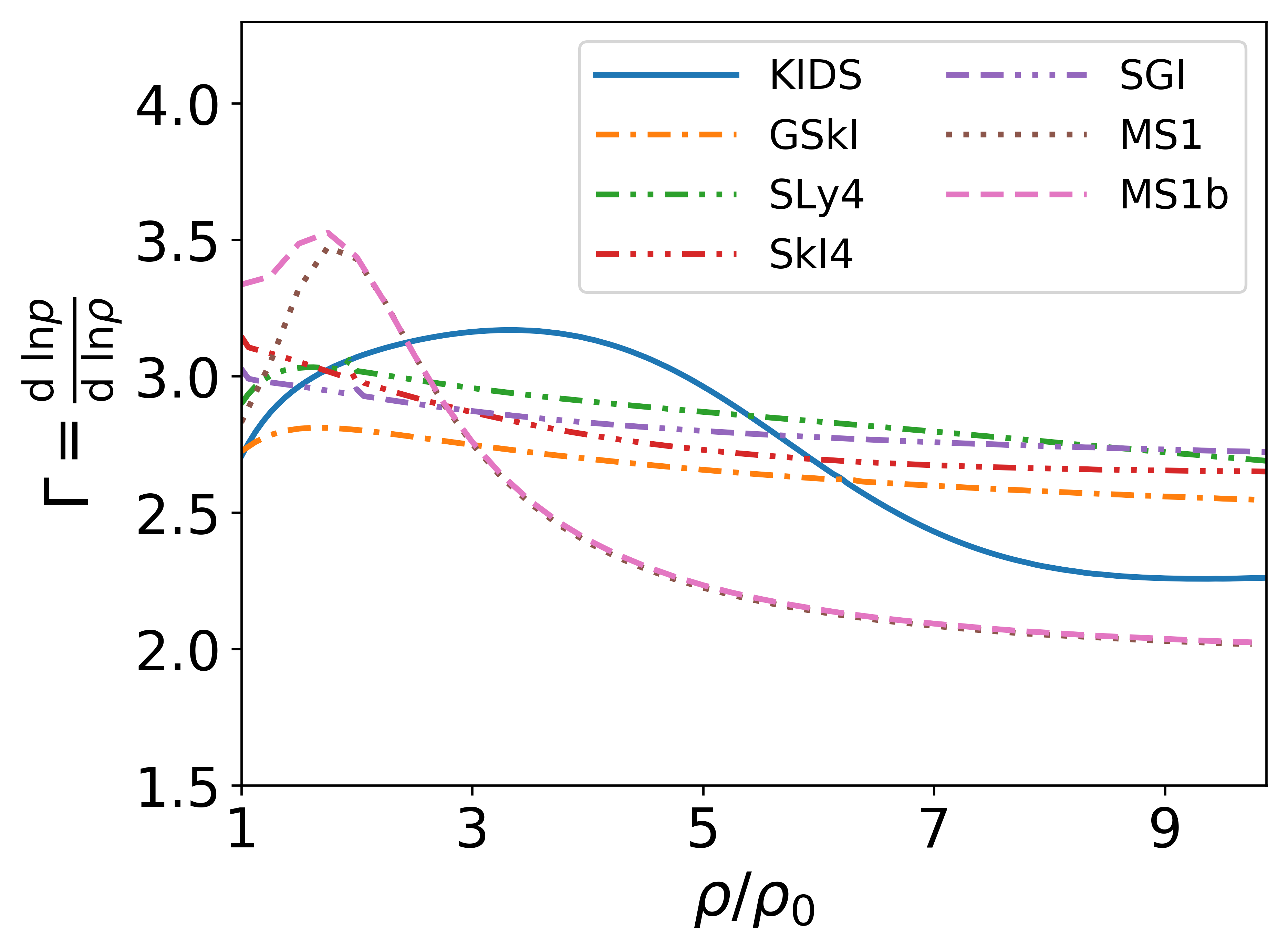}}
\centerline{(a) $\Gamma$ vs $\rho/\rho_0$}
\resizebox{0.4\textwidth}{!}{\includegraphics{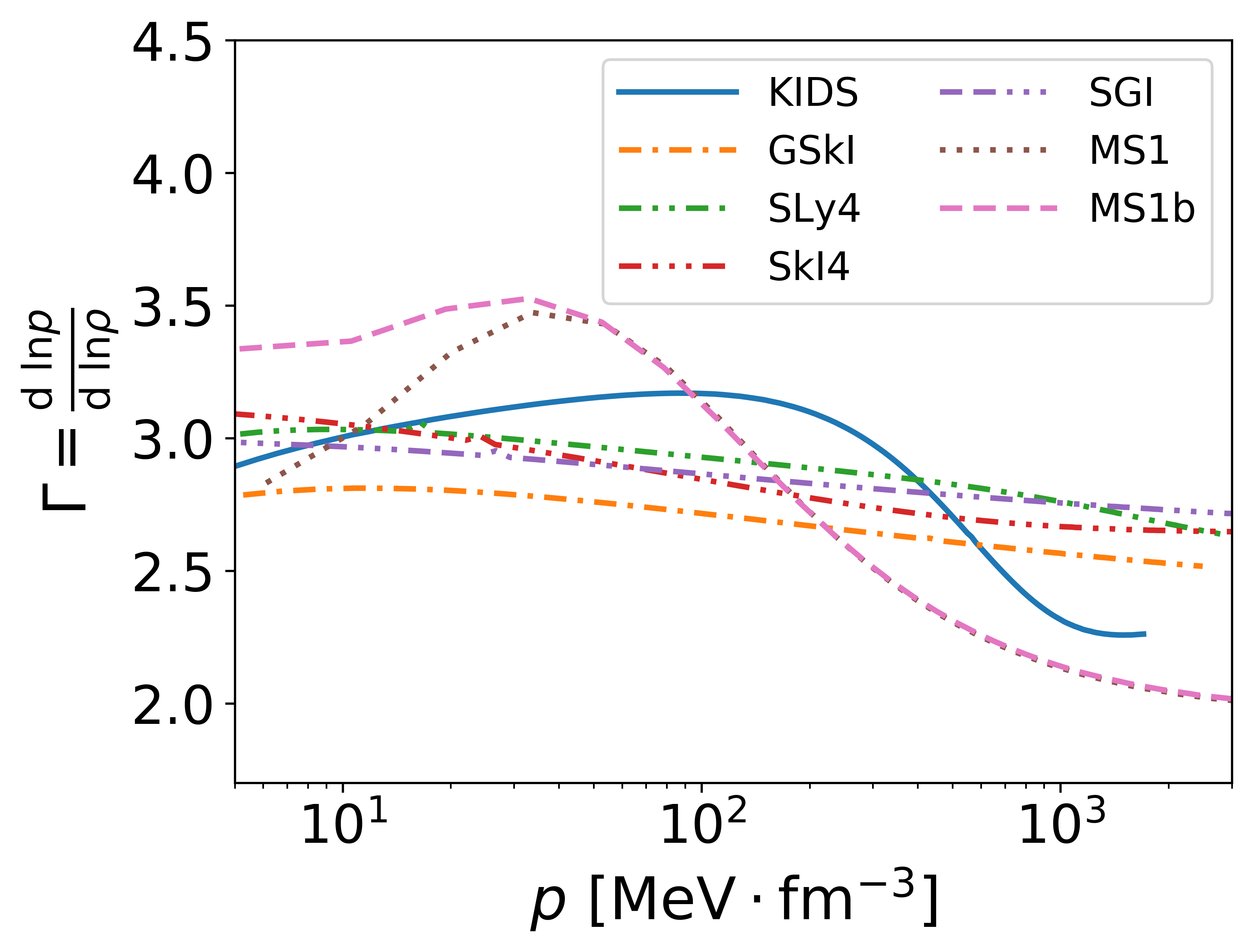}}
\centerline{(b) $\Gamma$ vs $p$}
\end{center}
\caption{Adabatic index $\Gamma$ vs density ($\rho/\rho_0$) and pressure ($p$) 
for various nuclear equations of state.} 
\label{fig3}
\end{figure}

\begin{figure}[t]
\begin{center}
\resizebox{0.4\textwidth}{!}{\includegraphics{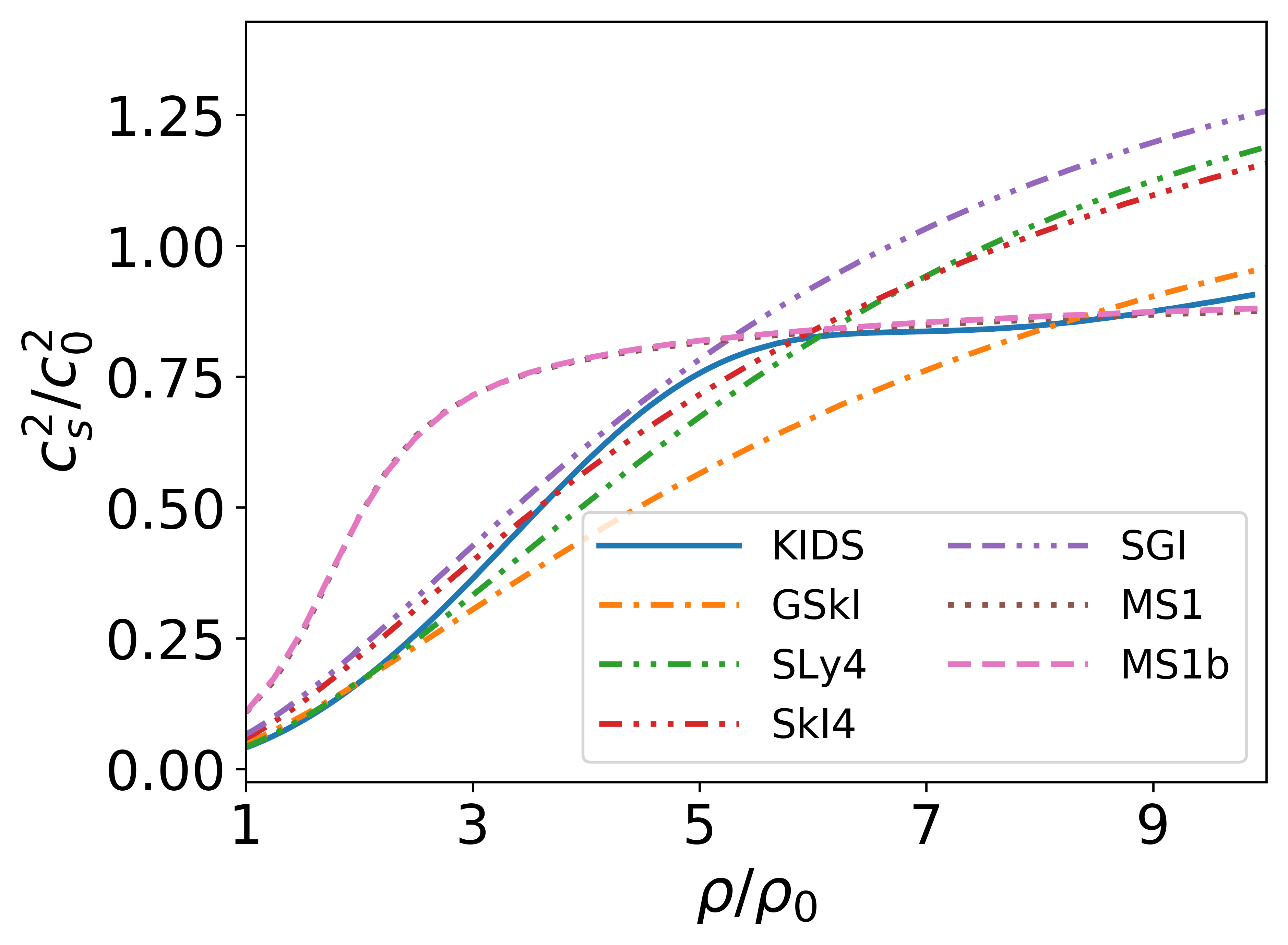}}
\centerline{(a) $c_c^2/c_0^2$ vs $\rho/\rho_0$ }
\centerline{\phantom{x}}
\resizebox{0.4\textwidth}{!}{\includegraphics{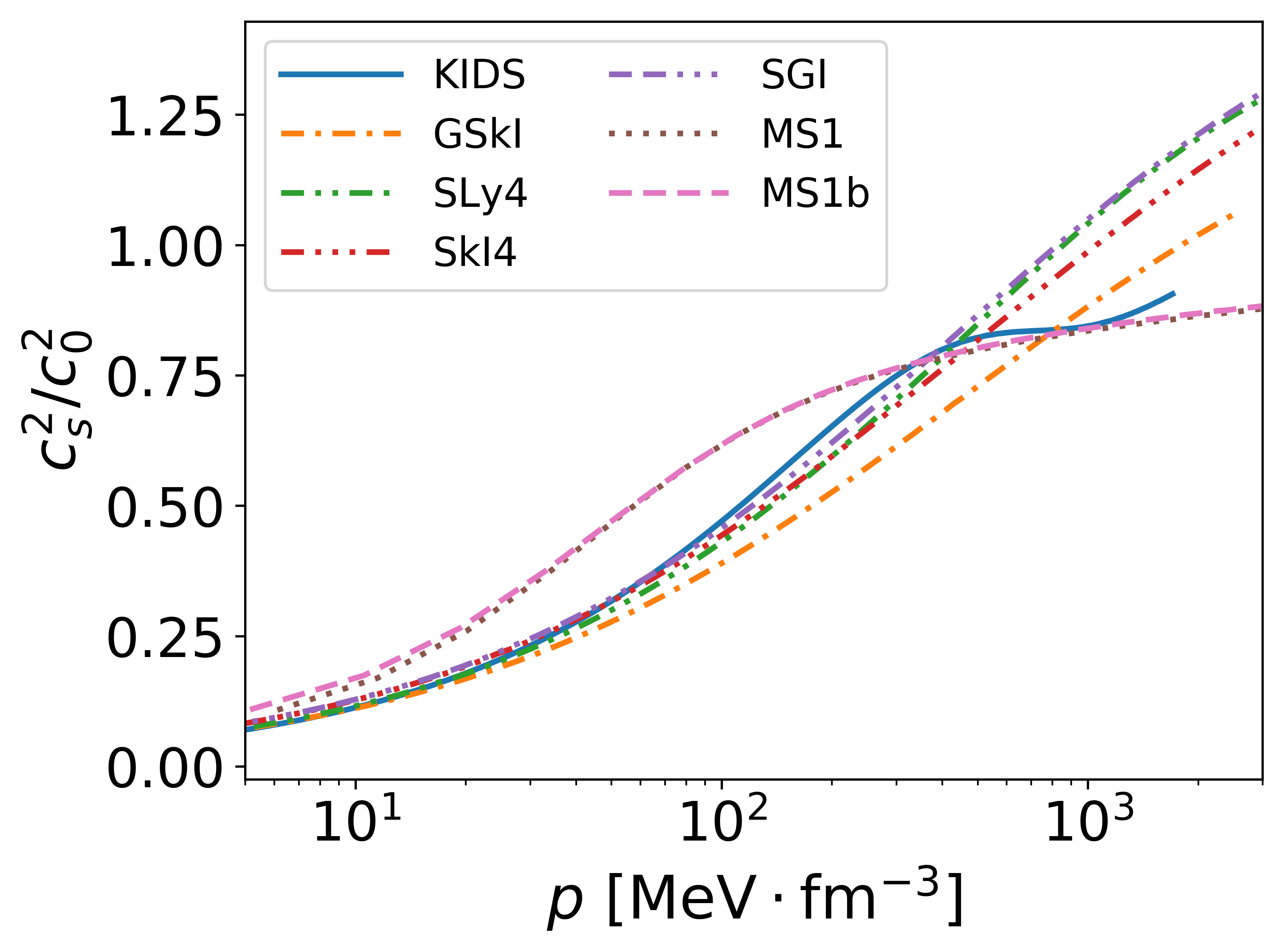}}
\centerline{(b) $c_s^2/c_0^2$ vs $p$ }
\end{center}
\caption{Speed of sound ($c_s$) for various equations of state~\cite{Kim2018a,Kim2018b}. $c_0$ is speed of light in vacuum.  }
\label{fig4}
\end{figure}

In Fig.~\ref{fig2}, the nuclear equations of state which we consider in this work are plotted. In order to understand the characteristics of each equation of state, we plot the adabatic index in Fig.~\ref{fig3};
\begin{equation}
\Gamma (p) \equiv  \frac{d (\ln p)}{d (\ln \rho)}. 
\end{equation}
An ideal gas equation of state can be characterized by a constant adiabatic index $\Gamma$, e.g., $\Gamma = 5/3$ for a non-relativistic ideal gas and $\Gamma=4/3$ for an ultra-relativistic ideal gas. However, as one can see in Fig.~\ref{fig3} (a), the adabatic index $\Gamma$ is not a constant for more realistic nuclear equations of state and $\Gamma > 2$ for most of the models at high densities because the strong interaction dominates. Recently, spectral expansions of the adiabatic index have been introduced to parameterize neutron star equations of state~\cite{Lindblom18} and used to analyze the tidal deformability of neutron stars~\cite{LSC18}. In Fig.~3 (b), the adiabatic index is plotted as a function of pressure for the comparison. Note that the adiabatic indexes of MS1 and MS1b at low (high) densities are much larger (smaller) than those of other models. On the other hand the adiabatic index of KIDS model shows distinctive behavior among the models that were considered in this work. In Fig.~\ref{fig4} speed of sound is plotted. At high density, even though KIDS satisfies the causality limit $c_s^2/c_0^2 < 1$, it gives large speed of sound, far above the ideal gas limit $c_s^2/c_0^2 = 1/3$. Note that large speed of sound at high density is required to make stiff equation of state which is consistent with current bservations~\cite{demorest2010,antoniadis2013,cromartie19}.

%Figure3

%Figure4

\begin{figure}[t]
\begin{center}
\resizebox{0.4\textwidth}{!}{\includegraphics{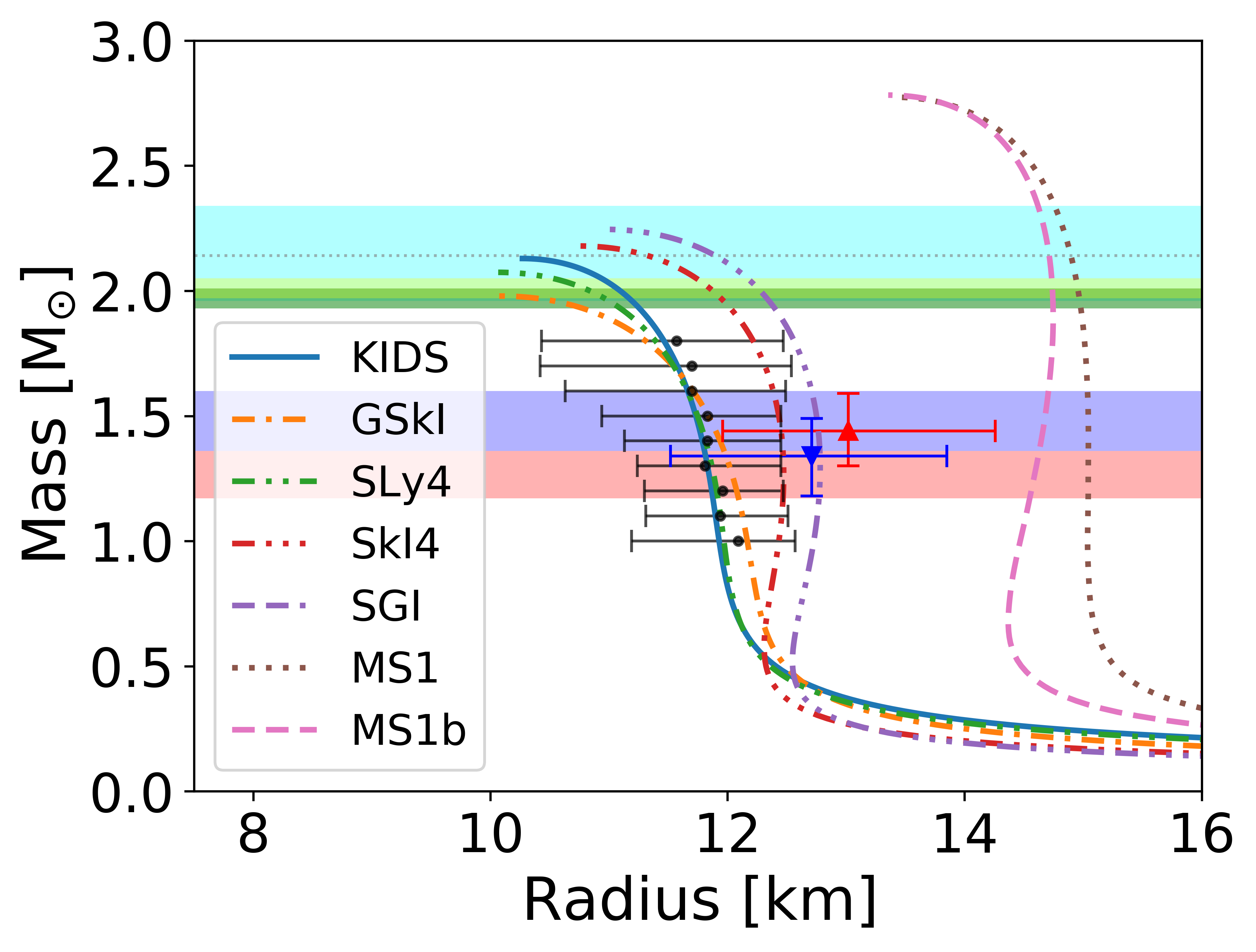}}
\centerline{(a) Mass vs Radius }
\resizebox{0.4\textwidth}{!}{\includegraphics{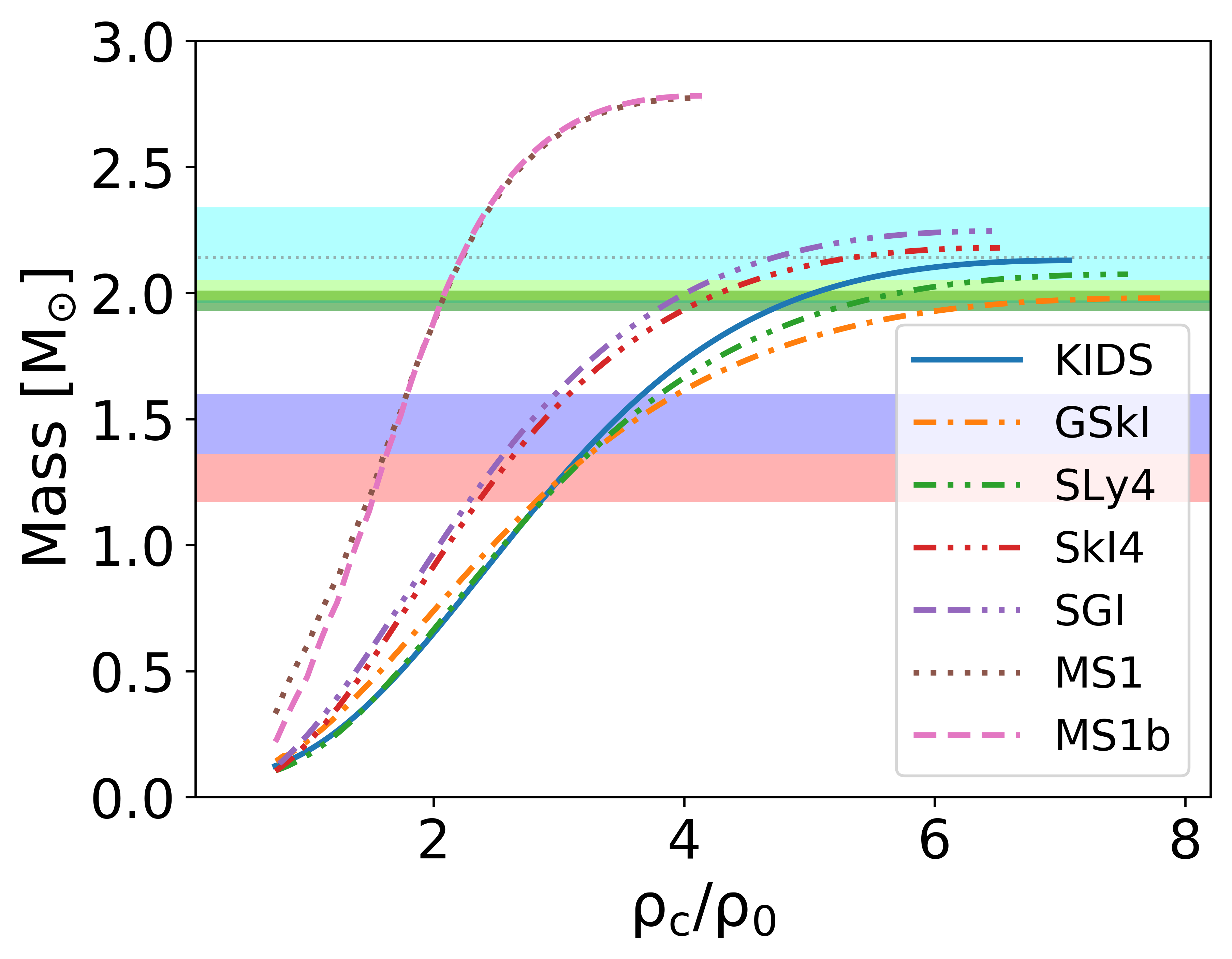}}
\centerline{(b) Mass vs Central density ($\rho_c$) }
\end{center}
\caption{ Mass vs radius and central density of neutron stars for various neutron star equations of state~\cite{Kim2018a,Kim2018b}. 
Upper three (cyan, yellow and green) horizontal bands indicate neutron star masses observed in white-dwarf neutron star binaries~\cite{demorest2010,antoniadis2013,cromartie19}.  Lower two (blue and red) horizontal bands indicate neutron star masses estimated in the neutron star binary merger GW170817~\cite{GW170817PRL}. Black circles with horizontal error bars around $R \sim 12$ km in (a) correspond to the probable radii of neutron stars estimated from X-ray bursts observation~\cite{steiner2010}. Blue and red triangles around $R \sim 13$ km in (a) correspond to the new estimates by NICER~\cite{Riley19,Miller19}.
}
\label{fig5}
\end{figure}

In Fig.~\ref{fig5}, masses, radii and central densities of neutron stars are summarized. As marked with cyan color, the most massive neutron star in neutron star-white dwarf binaries has $M= 2.14^{+0.20}_{-0.18} M_\odot$~\cite{cromartie19} which sets the lower limit of maximum mass of neutron stars. In Fig.~\ref{fig5} (a), radii of MS1 and MS1b are much larger than those of others for a given neutron star mass, 
and not consistent with current X-ray burst observations~\cite{steiner2010}. In Fig.~\ref{fig5} (b), for a given neutron star mass, central densities of MS1 and MS1b are much smaller than those of other models due to the larger adiabatic indexes (higher pressure) than other models at low densities as in Fig.~\ref{fig3}. Even though the adiabatic index of KIDS model is quite different from those of Skyrme force models (GSkI, SLy4, SkI4, SGI) as shown in Fig.~\ref{fig5} (a), masses and radii of neutron stars with KIDS model are less distinctive than the MS1 and MS1b models, and KIDS, GSkI and SLy4 models produce similar masses and radii which are consistent with current X-ray observations.

%Figure5

%Figure6
\begin{figure}[t]
\begin{center}
\resizebox{0.4\textwidth}{!}{\includegraphics{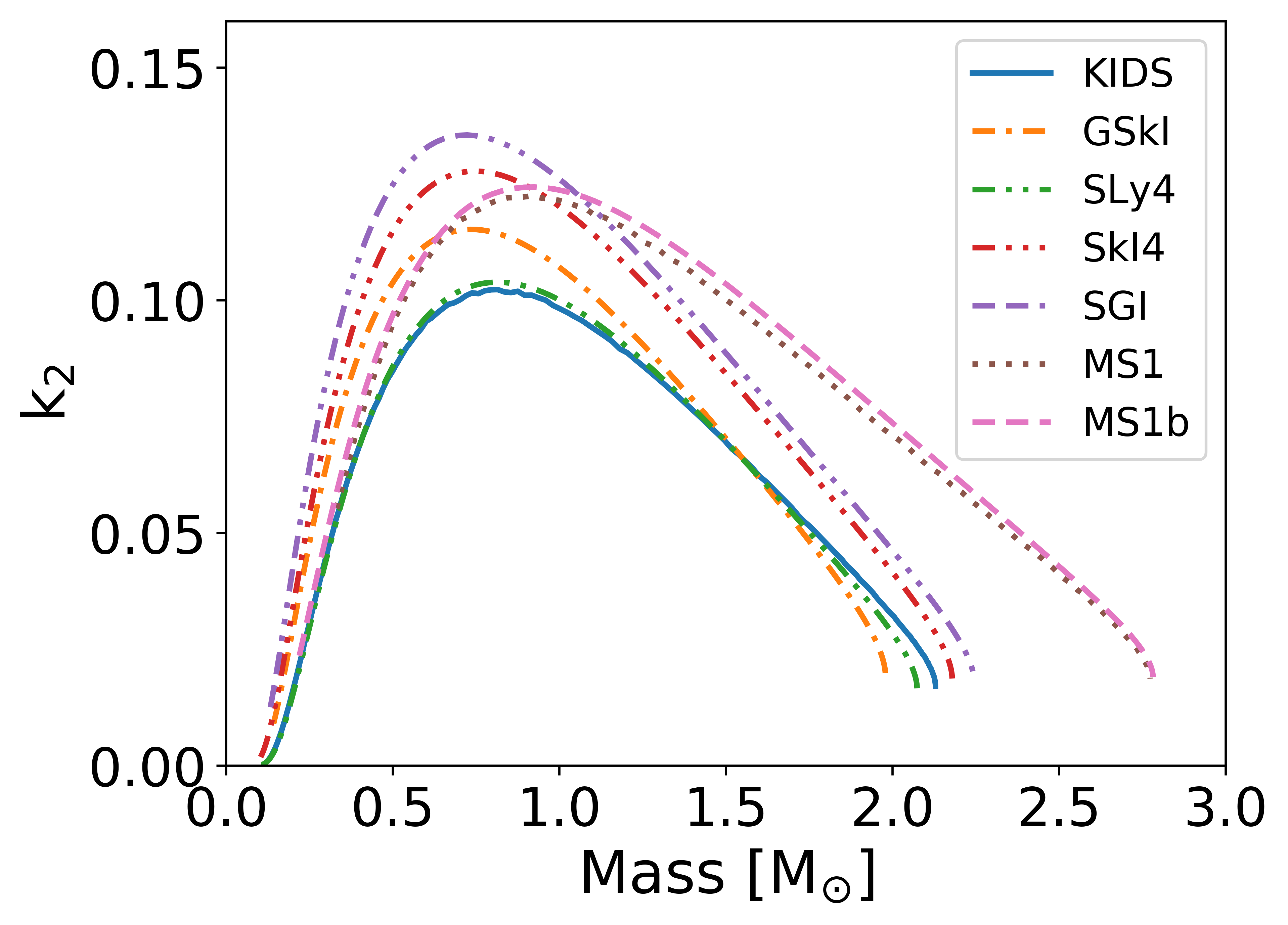}}    
\end{center}
\caption{Tidal Love number $(k_2)$ of a single neutron star~\cite{Kim2018a,Kim2018b}.} 
\label{fig6}
\end{figure}

%Figure7
\begin{figure}[t]
\begin{center}
\resizebox{0.39\textwidth}{!}{\includegraphics{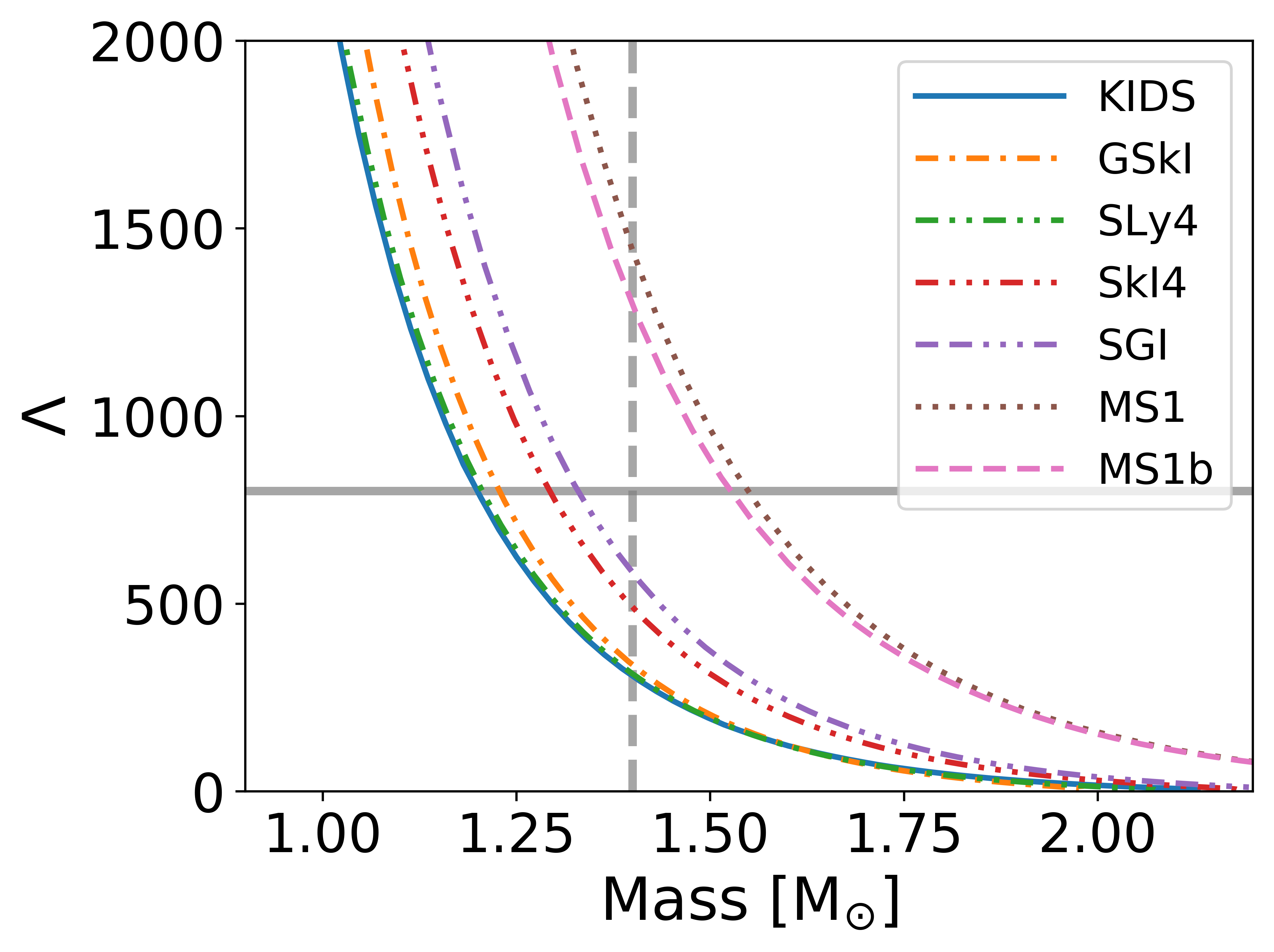}}   
\centerline{(a) $\Lambda$ vs Mass}
\resizebox{0.4\textwidth}{!}{\includegraphics{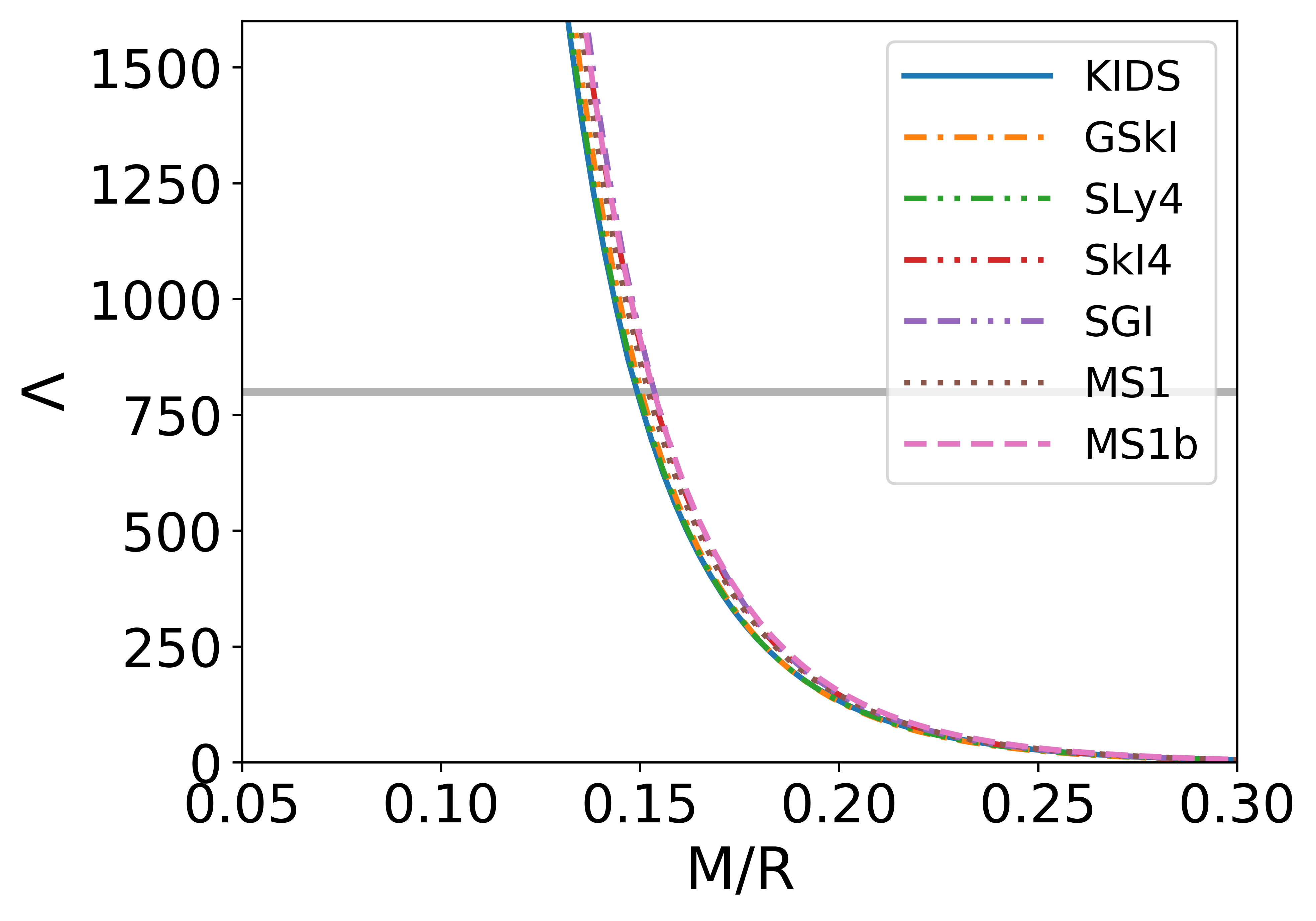}} 
\centerline{(b) $\Lambda$ vs Compactness ($M/R$)}
\end{center}
\caption{Dimensionless tidal deformability ($\Lambda$)  of a single neutron star~\cite{Kim2018a,Kim2018b}. Horizontal grey lines in all panels indicate the upper limit of $\Lambda$ with $M=1.4 M_{\odot}$ obtained from GW170817.
In (a), cross points of curves with the vertical dashed line correspond to $\Lambda$ at $M = 1.4 M_\odot$.}
\label{fig7}
\end{figure}

The dimensionless tidal deformability ($\Lambda$) which parameterizes the correlation between the external quadrupolar tidal field (${\cal E}_{ij}$) and the induced quadrupole moment of a neutron star ($Q_{ij}$) is defined as 
\begin{equation}
{\cal Q}_{ij} = - \Lambda\, M^5  {\cal E}_{ij},
\end{equation}
and the tidal Love number ($k_2$) can be obtained from the relation~\cite{GW170817PRL,Hin10,Fav14}
\begin{equation}
\Lambda = \frac{2}{3} \left(\frac{M}{R}\right)^{-5} k_2,
\label{eq4}
\end{equation}
where $M/R$ is called compactness of a neutron star. The dimensionless tidal deformability $\Lambda$ can be extracted from gravitational wave signals by accumulating the phase shifts of gravitational waves due to the $\Lambda$ contribution in gravitational waveform models~\cite{GW170817PRL,Fav14}. On the other hand, tidal Love number $k_2$ gives more intuitive information to understand the deformation of neutron stars. Tidal Love number of the neutron star is plotted in Fig.~\ref{fig6}. One can see that the tidal Love number converges to zero as the mass of a neutron star increases to its maximum value. This behavior could be understood by noting that the tidal deformability of a black hole is zero because there is no matter to be deformed. Tidal Love number also decreases toward zero as the mass of a neutron star decreases even though the radius increases because, for small mass neutron stars, density distribution is more important than the radius. Note that, as the neutron star mass decreases below 0.5 $M_\odot$, mass distribution is more centralized and the outer shell becomes more dilute.

In Fig.~\ref{fig7}, the dimensionless tidal deformability $\Lambda$ of a single neutron star is summarized for masses above 0.9 $M_\odot$ in which most of the observed neutron stars are distributed. In Fig.~\ref{fig7}(a), one can confirm that MS1 and MS1b are not consistent with the upper limit of the tidal deformability obtained from GW170817~\cite{GW170817PRL}. Other 5 models including KIDS are consistent with the current upper bound of the tidal deformability $\Lambda$. However, with more observations on the tidal deformability, one may be able to exclude some of the models because the largest difference in $\Lambda$ with $M=1.4 M_\odot$ is about a factor of two as in Fig.~\ref{fig7}(a). In Fig.~\ref{fig7}(b), $\Lambda$ is plotted as a function of compactness $M/R$ for masses above 0.9 $M_\odot$. In this mass range, $\Lambda$ mainly depends on the compactness independently of chosen models. This result can be understood from the results in Fig.~\ref{fig6} and Eq.~(\ref{eq4}). Since $k_2$ is almost linearly proportional to $(M/R)^{-1}$ for the masses above 0.9 $M_\odot$ as in Fig.~\ref{fig6}, the tidal deformability $\Lambda$ of a single neutron star becomes mainly a function of the compactness ($M/R$) in this mass range. Hence, from Eq.~(\ref{eq4}), one can get the approximate relation; $\Lambda \propto (M/R)^{-6}$.

In summary, we found that KIDS, GSkI and SLy4 models are consistent with both mass-radius constraints from X-ray observations and the upper bound of tidal deformability from gravitational-wave observations. MS1 and MS1b models are more or less excluded because their results are not consistent with masses and radii obtained from X-ray observations, nor with the upper bound of the tidal deformability obtained from gravitational-wave observations. The remaining two Skyrme force models (SkI4 and SGI) are marginally consistent with both observations because the radii are near the upper boundary of X-ray observations as in Fig.~\ref{fig5} and the tidal deformabilities are also close to the upper boundary of revised estimate of the tidal deformability, $\Lambda_{1.4} = 190^{+390}_{-120}$ of W170817~\cite{LSC18}. Hence, the validity of nuclear equations of state  can be further tested by future observations with less uncertainties.

%--------------------------------------------------------------------------------------------------
\section{Prospects} 
\label{sec4}

In this work, we show that the energy density functionals such as KIDS allow us to understand both finite nuclei and nuclear matter in a systematic way. Although we did not include the details in this review, the symmetry energy plays an important role in understanding the physical properties of the dense nuclear matter~\cite{Kra18,Rai19} and neutron stars~\cite{Lattimer14}. Producing exotic nuclei that have larger differences in the numbers of neutrons and protons and performing collision experiments using them will help us understand the symmetry energy at high densities. Such experiments will be available in the future or planned rare isotope facilities including RAON (Rare isotope Accelerator complex for ON-line experiments) in Korea~\cite{Ahn13}.

Recent measurement of a neutron star mass and radii by NICER~\cite{Riley19,Miller19} provides new constraints to the neutron star equations of state. The role of asymmetric nuclear matter properties, such as symmetry energy, has to be reanalyzed to satisfy the new observation. Energy density functionals including KIDS are very flexible and appropriate for the analysis because they have degrees of freedom to accommodate the new observation before fitting the nuclear properties. The work with KIDS in this direction is in progress.

In addition to GW170817, tidal deformability of neutron stars has been estimated from the neutron star binary merger GW190425; $\tilde\Lambda \le 600$ with low-spin prior and $\tilde\Lambda \le 1100$ with high-spin prior~\cite{GW190425}. Despite the uncertainties in the event rates of the neutron star-neutron star mergers, we could be still optimistic about the future discovery of the neutron star merger events and more accurate measurements of neutron star properties because the sensitivities of the gravitational wave detectors are expected to be improved.

\section*{Acknowledgements}

We were supported by the National Research Foundation of Korea (NRF) grant funded by the Korea government; 
Ministry of Science and ICT and Ministry of Education.
Y.M.K. NRF-2016R1A5A1013277 and NRF-2019R1C1C1010571;
% KK was supported by NRF grant funded by the Korea government (Ministry of Science and ICT and Ministry of Education) 
K.K. NRF-2016R1A5A1013277;\\
% CHH was supported by the NRF grant funded by the Korea government (MSIT) 
C.H.H. NRF-2018R1A5A1025563;\\
% CHL was supported by NRF grants funded by the Korea government  (Ministry of Sciencd and ICT and Ministry of Education)  
C.H.L. NRF-2016R1A5A1013277\\ and NRF-2018R1D1A1B07048599.

\end{document}